\documentclass[12pt]{article}
\usepackage{epsf}
\usepackage[dvips]{color}
\usepackage{exscale}
\usepackage[dvips]{graphicx}
\usepackage{amsmath}
\usepackage{amssymb}
\usepackage{latexsym}

\begin{document}
\setlength{\baselineskip}{18pt}
\begin{titlepage}

\vspace*{1.2cm}
\begin{center}
{\Large\bf GIMPs from Extra Dimensions}
\end{center}
\lineskip .75em
\vskip 1.5cm

\begin{center}
{\large 
Martin Holthausen\footnote[1]{E-mail:
\tt martin.holthausen@mpi-hd.mpg.de} and 
Ryo Takahashi\footnote[2]{E-mail: 
\tt ryo.takahashi@mpi-hd.mpg.de}
}\\

\vspace{1cm}

{\it Max-Planck-Institut f$\ddot{u}$r Kernphysik, Saupfercheckweg 1, 
69117 Heidelberg, Germany}\\

\vspace*{10mm}
{\bf Abstract}\\[5mm]
{\parbox{13cm}{\hspace{5mm}
We study a scalar field theory in a flat five-dimensional setup, where a scalar 
field lives in a bulk with a Dirichlet boundary condition, and give an 
 implementation of this setup to the Froggatt-Nielsen (FN) mechanism. It is 
shown that all couplings of physical field of the scalar with the all brane 
localized standard model particles are vanishing while realizing the usual FN 
mechanism. This setup gives the scalar a role as an only Gravitationally 
Interacting Massive Particle (GIMP), which is a candidate for dark matter.}}
\end{center}
\end{titlepage}
The Large Hadron Collider (LHC) experiments is just being started. One of the 
important missions of this big project is the discovery of the Higgs particle. 
The coupling to this particle is the origin of fermion masses in the standard 
model (SM) and it plays a crucial role in the electroweak (EW) symmetry 
breaking. The SM will be completed as a renormalizable theory when the Higgs is 
discovered. However, a theoretical problem still exists within the SM which is 
so-called the gauge hierarchy problem about the big desert between the EW and 
Planck scales. Theories with additional space dimensions
are interesting approaches towards solving this problem 
\cite{ArkaniHamed:1998rs,Randall:1999ee}. Generally, extra dimensional theories 
lead to rich phenomenologies, for instance, new heavy particles with masses of 
the compactification scale, the so-called Kaluza-Klein (KK) particles. In the 
Universal Extra Dimensions (UED) model \cite{ued}, the lightest KK particle with
 an odd parity is stable, and can be a candidate for the dark matter. The 
compactification scale is generally constrained to be larger than a few TeV by 
the EW precision measurements for the brane localized fermion scenario 
\cite{kk1}-\cite{kk5} but it can be  weakened to a few hundred GeV in the UED 
case due to the five-dimensional Lorentz symmetry \cite{ued,App,Gogo}. Another 
interesting phenomenological consequence is the top Yukawa deviation, which is a
 deviation of the Yukawa coupling between top and physical Higgs fields from 
naive SM expectation, induced from an existence of the brane localized Higgs 
potentials \cite{Haba:2009uu,Haba2} leading to Dirichlet type boundary 
conditions \cite{Haba:2009pb}, and the $SO(5)\times U(1)$ warped gauge-Higgs 
unification model \cite{hk}. The LHC experiment may give some suggestions for 
the TeV or few hundred GeV scale KK resonances and resultant phenomena induced 
from bulk fields.

On the other hand, the flavour problem, that is the origin of the three 
generations of the SM fermions and their Yukawa couplings, which determine the 
masses and mixings, is also one of the most important problems in the SM.
A fascinating approach to explain the flavour structure of the SM is to 
introduce some flavour symmetries broken at a high energy scale by an additional
 scalar (Higgs) field, the so-called flavon \cite{Froggatt:1978nt}. Such a field
 is introduced in a number of flavour models, e.g. for the purpose of realizing 
tri-bimaximal generation mixing \cite{Altarelli:2005yp} via non-abelian discrete
 flavour symmetries \cite{Hagedorn:2006ir}, etc..
 
In this letter, we study a scalar field theory in a five-dimensional setup, 
where a scalar field lives in a bulk with a Dirichlet boundary conditions. 
We then further discuss implementations of this setup to the Froggatt-Nielsen 
(FN) mechanism.

We consider a complex scalar field theory in a five-dimensional spacetime 
compactified on a flat line segment. The bulk-scalar kinetic action is given by
 \begin{eqnarray}
  S=-\int d^4x\int_{-L/2}^{+L/2}dz|\partial_M\Phi|^2, \label{action}
 \end{eqnarray}
where we write five-dimensional coordinates as $x^M=(x^\mu,z)$ with $\mu=0,1,2,3$
 and the extra dimension is compactified on a line segment $-L/2\leq z\leq L/2$.
 \footnote{One can consider in the usual extradimensional coordinate, $y$. In 
this case, the fundamental region becomes $0\leq y\leq L$, and $z$ is defined as
 $z\equiv y-L/2$.} Our metric convention is $(-++++)$. In a case of a free 
complex scalar field, $\Phi=(\Phi_R+i\Phi_I)/\sqrt{2}$, the variation of the 
action is given by
 \begin{eqnarray}
 \delta S = \int d^4x\int_{-L/2}^{+L/2}dz\bigg[\delta\Phi_X(\mathcal{P}\Phi_X) 
            +\delta(z-\frac{L}{2})\delta\Phi_X(-\partial_z\Phi_X)
            +\delta(z+\frac{L}{2})\delta\Phi_X(+\partial_z\Phi_X)\bigg],
 \end{eqnarray}
where $\mathcal{P}\equiv\Box+\partial_z^2$. The vacuum expectation value (VEV) 
of the scalar field, $\Phi^c$, is determined by the action principle, 
$\delta S=0$, that is, $\mathcal{P}\Phi_X^c=0$. Assuming unbroken 4D Lorentz 
invariance, the general solution of this equation (EOM) is $\Phi^c(z)=A+Bz$. The
 undetermined coefficients $A$ and $B$ can be fixed by taking boundary 
conditions (BCs) at $z=\pm L/2$. We obviously have four choices of combination 
of Dirichlet and Neumann BCs at $z=(-L/2,+L/2)$, namely, $(D,D)$, $(D,N)$, 
$(N,D)$, and $(N,N)$, where $D$ and $N$ denotes Dirichlet and Neumann BCs, 
respectively. These BCs are written as $\delta\Phi(x,z)|_{z=\xi}=0$ for the 
Dirichlet BC, and $\partial_z\Phi(x,z)|_{z=\xi}=0$ for the Neumann BC, where 
$\xi$ is taken as $+L/2$ or $-L/2$ in each case. A different choice of BCs 
corresponds to a different choice of the theory. The theory is fixed once one 
chooses one of the four conditions. In this letter, we suppose that all SM 
particles are localized on $z=+L/2$ brane and focus on a Dirichlet BC at the 
brane. Therefore, the discussed BCs are restricted to $(D,D)$ or $(N,D)$. The 
most general BCs for each case can be written as
 \begin{eqnarray}
  &&(\Phi(x,z)|_{z=-L/2},\Phi(x,z)|_{z=+L/2})=(v_-,v_+), \label{DD}\\
  &&(\partial_z\Phi(x,z)|_{z=-L/2},\Phi(x,z)|_{z=+L/2})=(0,v_+), \label{ND}
 \end{eqnarray}
for $(D,D)$ and $(N,D)$, respectively, where $v_-$ and $v_+$ are constants of 
mass dimension $[3/2]$. The BCs \eqref{DD} and \eqref{ND} fix the VEV to be the 
value $v_+$ on $z=+L/2$ brane, while requiring the quantum fluctuation 
to be vanishing at the boundary. These BCs also determine the coefficients $A$ 
and $B$, that is, the VEV profile in the extra-dimensional direction as
 \begin{eqnarray}
  \Phi^c(z)=
   \left\{
    \begin{array}{lll}
     \frac{v_++v_-}{2}+\frac{v_+-v_-}{L}z & \mbox{ for } & (D,D), \\
     v_+                                 & \mbox{ for } & (N,D).
    \end{array}
   \right. \label{VEVp}
 \end{eqnarray}
It is easily seen that the resultant VEV profile for $(N,D)$ case becomes flat 
in the extra dimension due to the Neumann BC at the $z=-L/2$ brane while the one
 for $(D,D)$ case linearly depends on the extradimensional coordinate. In a case
 of $v_-=v_+$, the profile for $(D,D)$ BCs also becomes flat, $\Phi^c(z)=v_+$. 

\begin{figure}
\begin{center}
 \includegraphics[scale = 0.83]{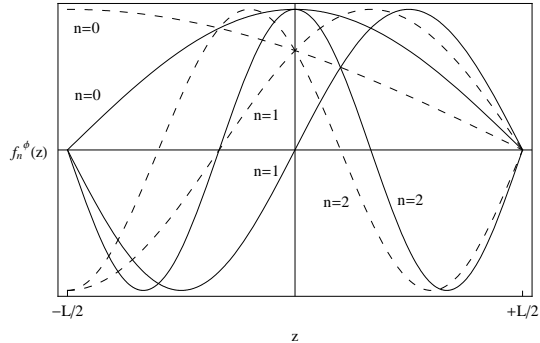}
\end{center}\vspace{-5mm}

\caption{The wave function profiles of $n=0,1$, and $2$ modes in \eqref{waveDD} 
and \eqref{waveND}: The solid and dashed curves correspond to the $(D,D)$ and 
$(N,D)$ cases, respectively.}
\label{fig1}
\end{figure}

Next, let us consider the profile of quantum fluctuation of the scalar field. 
We utilize the background field method, separating the field into the classical 
field and quantum fluctuation:
 \begin{eqnarray}
  \Phi(x,z)=\Phi^c(z)+\frac{1}{\sqrt{2}}\left[\phi(x,z)+i\chi(x,z)\right].
  \label{sepa}
 \end{eqnarray}
We put separation \eqref{sepa} into \eqref{action} and expand up to the 
quadratic terms of the field $\phi$ as\footnote{The same expansion is taken for 
$\chi(x,z)$. See e.g. Appendix B in \cite{Haba:2009uu} for the derivation of 
action.} 
 \begin{eqnarray}
  S_\phi= \int d^4x\int_{-L/2}^{+L/2}dz
           \bigg[\frac{1}{2}\phi\left(\Box+\partial_z^2\right)\phi 
        +\frac{\delta(z-L/2)}{2}\phi(-\partial_z\phi)
            +\frac{\delta(z+L/2)}{2}\phi(\partial_z\phi)\bigg].
  \label{phi-ac}
 \end{eqnarray} 
Hereafter we focus on only $\phi(x,z)$ field for our purpose. The KK expansion 
for $\phi(x,z)$ is given by
 \begin{eqnarray}
  \phi(x,z)=\sum_{n=0}^\infty f_n^\phi(z)\phi_n(x), \label{KK}
 \end{eqnarray}
where $f_n(z)$ are eigenfunction of differential operator in the free action 
\eqref{phi-ac}:
 \begin{eqnarray}
  \partial_z^2f_n^\phi(z)=-\mu_{\phi n}^2f_n^\phi(z).
 \end{eqnarray}
The general 
solution of this equation for each $n$th mode is written as 
$f_n^\phi(z)=\alpha_n\cos(\mu_{\phi n}z)+\beta_n\sin(\mu_{\phi n}z)$. 
In total there are now three unknown constants for each $n$th mode, $\alpha_n$, 
$\beta_n$, and $\mu_{\phi n}$. Two of the three are fixed by the two BCs at 
$z=\pm L/2$ while the 
last one is fixed by the normalization 
$\int_{-L/2}^{+L/2}dzf_n^\phi(z)f_m^\phi(z)=\delta_{nm}$. In the following we are focusing 
on two specific choices of BCs, namely the $(D,D)$ and $(N,D)$ cases. Then, the EOM under 
such BCs determines the VEV profiles given in \eqref{VEVp} and expansion 
\eqref{sepa} leads to the following BCs for quantum fluctuation (substituting 
both \eqref{VEVp} and \eqref{sepa} to \eqref{DD} and \eqref{ND}),
 \begin{eqnarray}
  &&(f_n^\phi(z)|_{z=-L/2},f_n^\phi(z)|_{z=+L/2})=0, \\
  &&(\partial_zf_n^\phi(z)|_{z=-L/2},f_n^\phi(z)|_{z=+L/2})=0,
 \end{eqnarray}
 for $(D,D)$ and $(N,D)$, respectively. Therefore, we can obtain 
the wave function profile of the quantum fluctuations as
 \begin{eqnarray}
  f_n^\phi(z) &=&
   \left\{
    \begin{array}{lll}
     \sqrt{\frac{2}{L}}\cos\left(\frac{(n+1)\pi}{L}z\right) & \mbox{ for even} & n, \\
     \sqrt{\frac{2}{L}}\sin\left(\frac{(n+1)\pi}{L}z\right) & \mbox{ for odd} 
& n,
    \end{array}
   \right. \label{waveDD} \\
  f_n^\phi(z) &=& \sqrt{\frac{1}{L}}\left[\cos\left(\frac{(2n+1)\pi}{2L}z\right)
                 -(-1)^n\sin\left(\frac{(2n+1)\pi}{2L}z\right)\right]. 
                 \label{waveND}
 \end{eqnarray}
for $(D,D)$ and $(N,D)$ BCs, respectively. The wave function profiles of 
$n=0,1$, and $2$ modes in the extra-dimension are shown in FIG. \ref{fig1}. 
They mean that a flat zero-mode profile in the Neumann BC case is deformed to 
the cosine function of $f_0^\phi(z)=\sqrt{2/L}\cos(\pi z/L)$ through the 
Dirichlet BCs at $z=\pm L/2$ for the $(D,D)$ case. The profiles are described by
 a combination of the sine and cosine functions due to the Neumann BC at 
$z=-L/2$ and the Dirichlet at $z=+L/2$ for the $(N,D)$ case.

The $n$th scalar mass is calculated to be
 \begin{eqnarray}
  m_{\phi_n}^2=
   \left\{
    \begin{array}{ll}
     \left(\frac{(n+1)\pi}{L}\right)^2 & \mbox{ for }(D,D), \\
     \left(\frac{(2n+1)\pi}{2L}\right)^2 & \mbox{ for }(N,D).
    \end{array}
   \right.
 \end{eqnarray}
which shows the lowest ($n=0$) mode has a KK mass, $m_{\mbox{{\scriptsize KK}}} 
\equiv\pi/L$, for the $(D,D)$ case. On the other hand, the mass of the lowest 
mode for the $(N,D)$ case becomes a half of KK mass, $m_{\mbox{{\scriptsize KK}}}/2$. 
These features are just results from the Dirichlet BC, that is, the lowest mode 
mass is pushed up to the KK mass in the case of Dirichlet BCs at both 
boundaries, and the mass is pushed up to only a half of KK mass when the 
Dirichlet BC is taken at one boundary. The above discussions can be similarly 
applied to the $\chi$ field. The important point of this type of setup is that 
with the Dirichlet BC(s) the wave function profile of the physical field is 
vanishing at the boundary while the VEV of the scalar field can be obtained due 
to the Dirichlet BC without contradiction to the action principle. The mass 
of this physical field is of the order of the compactification scale as massless
 modes are forbidden by the BCs. 

Next, we propose an implementation of this setup. We investigate an 
identification of the bulk scalar field with a flavon. The origin of the three 
generations of the SM fermions and their Yukawa couplings determining the masses
 and mixings is one of the most important problems in particle physics. An 
introduction of family symmetries is a common approach in order to explain the 
origin of fermion masses and mixings. Such symmetries must be broken at a high 
energy scale, and additional scalar fields, the so-called flavons, are required
to break the symmetries. Then effective Yukawa couplings can be induced from 
non-renormalizable operators, which generates hierarchical Yukawa structures 
through the Froggatt-Nielsen (FN) mechanism \cite{Froggatt:1978nt}.

In the FN mechanism, scalar fields (flavons) are introduced that 
are charged under a family symmetry which acts on the different generations of 
SM fermions. Once the symmetry is broken at a high energy scale, the effective 
Yukawa coupling can be determined by the charges of fields and the VEV of 
flavons. Here, we investigate a case that the flavons are bulk scalar fields 
described by the above setup \eqref{action} with the Dirichlet BC \eqref{DD} or 
\eqref{ND} and the SM fermions are localized at the $z=+L/2$ brane. When we 
introduce a abelian family symmetry and one flavon for simplicity, the effective
 mass term for the SM fermions in four dimensions is given by
 \begin{eqnarray}
  \int_{-L/2}^{+L/2}dz\delta(z-\frac{L}{2})c_{ij}
               \left(\frac{\Phi}{\Lambda^{3/2}}\right)^{N_{ij}}\bar{F}_{Li}F_{Rj}H
               +h.c.,
 \end{eqnarray}
where $c_{ij}$ is a dimensionless coupling of order one and $N_{ij}$ are 
determined by the charges of flavon and SM fermions. After expanding $\Phi$ as 
in \eqref{sepa} and integrating the five dimensional direction, we obtain
 \begin{eqnarray}
  c_{ij}\epsilon^{N_{ij}}\bar{F}_{Li}F_{Rj}H+h.c., 
 \end{eqnarray}
where $\epsilon\equiv v_+/\Lambda^{3/2}$. This is the usual result of the FN 
mechanism except for the different mass dimension of the VEV in $\epsilon$. 
However, notice that the lowest modes of the physical fields $\phi_0(x)$ in the 
KK expansion \eqref{KK} certainly exist with the KK (a half of KK) mass induced 
from the bulk kinetic term \eqref{action} with $(D(N),D)$ BC. The physical 
states do not have any couplings with the brane localized SM fields because the 
wave function profiles are vanishing at boundary while the VEV can be obtained 
due to the Dirichlet BC. Therefore, the lowest mode of the flavon remains as a 
stable particle. The physical state of flavon becomes only Gravitationally 
Interacting Massive Particle (GIMP) which can be a candidate for the dark 
matter.

One should note, however, that the stability of the  dark matter candidate would
 be spoilt by the introduction of terms of the type 
 \begin{eqnarray}
  \int d^5x \delta(z-L/2)(\partial_5 \phi)\times \mathrm{SM \ fields}
 \end{eqnarray}
which would couple the KK modes to the SM fields.  Any bulk interactions will 
generally induce such disastrous brane interactions \cite{Georgi:2000ks}. The 
interpretation of the lightest KK mode of the scalar as the dark matter particle
 therefore requires that there must not be any bulk interactions of the scalar, 
meaning a free theory in the bulk. It is clear that since the KK sector is 
completely decoupled from the SM, there will be no interactions between the two
sectors induced by quantum corrections and the lightest KK mode is therefore 
stable. The stability is due to a 'superselection rule' between the two 
decoupled sectors whose decoupling is a consequence of the Dirichlet BCs even in
 the presence of Yukawa-type brane interactions. The relevant observation in 
this paper is thus the fact that the brane interactions might lead to a 
non-vanishing VEV of the field, which together with the Froggatt-Nielsen type 
interactions will lead to an explanation of the fermion mass hierarchy without 
the introduction of  a FN field in the 4D spectrum of the theory. The stable 
lightest KK mode is an only gravitationally interacting dark matter candidate. 

In the above implementation of Dirichlet BCs, the family symmetry is broken by 
the BC. The VEV related with the symmetry breaking scale and the cutoff scale 
could be taken as arbitrary high energy scales as long as $\epsilon\ll1$ is 
satisfied. The variant of the FN mechanism discussed above is only one example 
of this setup with the bulk flavon with Dirichlet BC and it can be applied to 
practically any flavour models with non-SM Higgs fields 
\cite{Altarelli:2005yp,Hagedorn:2006ir}. The implementation gives the fields a 
role as the GIMP. In some cases, constraints on flavour models from the 
electroweak precision measurements would be relaxed because the wave functions 
of non-SM bulk Higgs fields vanish at boundary. 

Finally, we comment on other extra-dimensional backgrounds which make it 
possible to identify our stable particle as a dark matter candidate. Our  
stabilization was based on a flat five-dimensional spacetime. If we extend the 
setup with the Dirichlet BC(s) to a larger number of extra-dimensions such as a 
six-dimensional model \cite{ArkaniHamed:1998rs} or to a warped extra-dimension 
model \cite{Randall:1999ee}, the hierarchy problem can be solved.  A realistic 
model, which does not suffer from the hierarchy problem, must be constructed on 
such backgrounds in studies of extra-dimensions. To get a viable DM candidate 
from our stabilization, we also have to extend the mechanism to a model with 
larger number of extra-dimension. Our KK-flavon, which has only gravitational
interactions, has survived until the present epoch, and its energy density can 
dominate but must not exceed the present DM one. Thus, we need 
$\Omega_\phi\leq\Omega_{\text{DM}},$ where $\Omega_\phi$ and $\Omega_{\text{DM}}$ are 
density parameters of KK-flavon and DM, respectively. The interaction rate of 
the KK-flavon can be roughly estimated as $\Gamma_\phi\sim T^5/M_{\text{pl}}^4$ with
 four-dimensional Planck mass, $M_{\text{pl}}$. The decoupling and non-exceeding 
conditions for KK-flavon constrain the KK-flavon mass to be smaller than 
$\mathcal{O}$(keV), $m_\phi\sim\Omega_\phi 
h^2g~(4.4\mbox{eV})\lesssim\mathcal{O}$(keV), where $h$ and $g$ are the current
dimensionless Hubble constant and degrees of freedom of order a few hundred at 
the decoupling temperature. This means that we require a larger number of 
extra-dimensions\footnote{Notice that the KK scale in $4+\delta$ dimensional 
corresponds to $m_\phi\sim\mathcal{O}(10^{-1})$meV, $\mathcal{O}$(10)keV, 
$\mathcal{O}$(10)MeV for $\delta=2,4,6$ with $4+\delta$ dimensional Planck mass 
of order $\mathcal{O}$(TeV), respectively.} in order to obtain a correct energy 
density of KK-flavons and to solve the gauge hierarchy problem without 
contradiction with the current cosmological observations and experimental limits
 for a deviation of the Newton's law. More detailed phenomenological predictions
 of this scenario should be compared with a DM model of sterile (lightest 
right-handed) neutrino with keV mass scale, and such discussions will be given 
in a separate publication.\\

 We have studied a scalar field theory in a flat five-dimensional setup with the
 Dirichlet type BC at $z=+L/2$ brane. The wave-function profiles of the physical
 field are deformed by the Dirichlet BC(s) in the setup. As the results, the 
physical field profiles vanish at the brane while a finite VEV can generally be 
obtained without contradiction to the action principle. The lowest mode 
masses of these fields are pushed up to a KK (a half of KK) scale by the 
$(D(N),D)$ BC. 

We have also proposed an implementation of this setup to flavon physics. The 
bulk scalar and all SM fields have been assumed to be a flavon in the FN 
mechanism and brane localized fields, respectively. 
It has been shown
that all couplings of the physical field of the flavon with the SM
particles are vanishing due to the Dirichlet BC while realizing the usual
FN mechanism in the free theory on the bulk. 
However, the physical field of the flavon with a KK mass induced from the bulk 
kinetic term certainly exist in the theory. This setup gives the flavon a role 
as an only Gravitationally Interacting Massive Particle (GIMP), which is a 
candidate for dark matter. This mechanism can be implemented in a number of 
flavor models with non-SM Higgs fields apart from the FN mechanism presented 
here.

\subsection*{Acknowledgments}
This work is supported by the DFG-SFB TR 27.


\begin{thebibliography}{99}
\bibitem{ArkaniHamed:1998rs}
 N.~Arkani-Hamed, S.~Dimopoulos and G.~R.~Dvali,
 Phys.\ Lett.\  B {\bf 429} (1998) 263,
 I.~Antoniadis, N.~Arkani-Hamed, S.~Dimopoulos and G.~R.~Dvali,
 Phys.\ Lett.\  B {\bf 436} (1998) 257.

\bibitem{Randall:1999ee}
 L.~Randall and R.~Sundrum,
 Phys.\ Rev.\ Lett.\  {\bf 83} (1999) 3370.

\bibitem{ued}
  T.~Appelquist, H.~C.~Cheng and B.~A.~Dobrescu,
  Phys.\ Rev.\  D {\bf 64} (2001) 035002.

\bibitem{kk1}
  P.~Nath and M.~Yamaguchi,
  Phys.\ Rev.\  D {\bf 60}, 116004 (1999).

\bibitem{kk2}
  M.~Masip and A.~Pomarol,
  Phys.\ Rev.\  D {\bf 60}, 096005 (1999).

\bibitem{kk3}
  T.~G.~Rizzo and J.~D.~Wells,
  Phys.\ Rev.\  D {\bf 61}, 016007 (2000).

\bibitem{kk4}  
  A.~Strumia,
  Phys.\ Lett.\  B {\bf 466}, 107 (1999).

\bibitem{kk5}  
  C.~D.~Carone,
  Phys.\ Rev.\  D {\bf 61}, 015008 (2000).

\bibitem{App}
 T.~Appelquist and H.~U.~Yee,
 Phys.\ Rev.\  D {\bf 67} (2003) 055002. 

\bibitem{Gogo}
 I.~Gogoladze and C.~Macesanu,
 Phys.\ Rev.\  D {\bf 74} (2006) 093012. 

\bibitem{Haba:2009uu}
  N.~Haba, K.~Oda and R.~Takahashi,
  Nucl.\ Phys.\  B {\bf 821} (2009) 74
  [Erratum-ibid.\  {\bf 824} (2010) 331].

\bibitem{Haba2}
  N.~Haba, K.~Oda and R.~Takahashi,
  arXiv:0910.4528 [hep-ph].

\bibitem{Haba:2009pb}
  N.~Haba, K.~Oda and R.~Takahashi,
  arXiv:0910.3356 [hep-ph].

\bibitem{hk}  
  Y.~Hosotani and Y.~Kobayashi,
  Phys.\ Lett.\  B {\bf 674} (2009) 192.

\bibitem{Froggatt:1978nt}
  C.~D.~Froggatt and H.~B.~Nielsen,
  Nucl.\ Phys.\  B {\bf 147} (1979) 277.

\bibitem{Altarelli:2005yp}
  G.~Altarelli and F.~Feruglio,
  Nucl.\ Phys.\  B {\bf 720} (2005) 64;
  S.~F.~King,
  JHEP {\bf 0508}, 105 (2005);
  I.~de Medeiros Varzielas and G.~G.~Ross,
  Nucl.\ Phys.\  B {\bf 733} (2006) 31;
  E.~Ma,
  Phys.\ Lett.\  B {\bf 632} (2006) 352;
  A.~Zee,
  Phys.\ Lett.\  B {\bf 630} (2005) 58;
  W.~Grimus and L.~Lavoura,
  JHEP {\bf 0601} (2006) 018;
  E.~Ma,
  Phys.\ Rev.\  D {\bf 73} (2006) 057304;
  G.~Altarelli and F.~Feruglio,
  Nucl.\ Phys.\  B {\bf 741}, 215 (2006);
  J.~E.~Kim and J.~C.~Park,
  JHEP {\bf 0605} (2006) 017;
  I.~de Medeiros Varzielas, S.~F.~King and G.~G.~Ross,
  Phys.\ Lett.\  B {\bf 644}, 153 (2007);
  R.~N.~Mohapatra, S.~Nasri and H.~B.~Yu,
  Phys.\ Lett.\  B {\bf 639} (2006) 318;
  I.~de Medeiros Varzielas, S.~F.~King and G.~G.~Ross,
  Phys.\ Lett.\  B {\bf 648} (2007) 201;
  E.~Ma,
  Mod.\ Phys.\ Lett.\  A {\bf 21}, 2931 (2006);
  G.~Altarelli, F.~Feruglio and Y.~Lin,
  Nucl.\ Phys.\  B {\bf 775} (2007) 31;
  H.~Zhang,
  Phys.\ Lett.\  B {\bf 655} (2007) 132;
  P.~D.~Carr and P.~H.~Frampton,
  [arXiv:hep-ph/0701034];
  M.~C.~Chen and K.~T.~Mahanthappa,
  Phys.\ Lett.\  B {\bf 652} (2007) 34;
  C.~Luhn, S.~Nasri and P.~Ramond,
  Phys.\ Lett.\  B {\bf 652} (2007) 27;
  Y.~Koide,
  arXiv:0707.0899 [hep-ph];
  E.~Ma,
  Phys.\ Lett.\  B {\bf 660}, 505 (2008);
  F.~Bazzocchi, S.~Morisi and M.~Picariello,
  Phys.\ Lett.\  B {\bf 659} (2008) 628;
  F.~Plentinger, G.~Seidl and W.~Winter,
  JHEP {\bf 0804} (2008) 077;
  F.~Plentinger and G.~Seidl,
  Phys.\ Rev.\  D {\bf 78} (2008) 045004;
  S.~Antusch, S.~F.~King and M.~Malinsky,
  JHEP {\bf 0805} (2008) 066;
  Y.~Lin,
  Nucl.\ Phys.\  B {\bf 813} (2009) 91;
  N.~Haba, R.~Takahashi, M.~Tanimoto and K.~Yoshioka,
  Phys.\ Rev.\  D {\bf 78} (2008) 113002.

\bibitem{Hagedorn:2006ir}
  C.~Hagedorn, M.~Lindner and F.~Plentinger,
  Phys.\ Rev.\  D {\bf 74} (2006) 025007;
  S.~Kaneko, H.~Sawanaka, T.~Shingai, M.~Tanimoto and K.~Yoshioka,
  Prog.\ Theor.\ Phys.\  {\bf 117} (2007) 161;
  S.~F.~King and M.~Malinsky,
  Phys.\ Lett.\  B {\bf 645} (2007) 351;
  S.~Morisi, M.~Picariello and E.~Torrente-Lujan,
  Phys.\ Rev.\  D {\bf 75} (2007) 075015;
  F.~Feruglio, C.~Hagedorn, Y.~Lin and L.~Merlo,
  Nucl.\ Phys.\  B {\bf 775} (2007) 120;
  F.~Yin,
  Phys.\ Rev.\  D {\bf 75} (2007) 073010;
  W.~Grimus and H.~Kuhbock,
  Phys.\ Rev.\  D {\bf 77}, 055008 (2008);
  B.~Brahmachari, S.~Choubey and M.~Mitra,
  Phys.\ Rev.\  D {\bf 77} (2008) 073008
  [Erratum-ibid.\  D {\bf 77} (2008) 119901];
  H.~Ishimori, T.~Kobayashi, H.~Ohki, Y.~Omura, R.~Takahashi and M.~Tanimoto,
  Phys.\ Lett.\  B {\bf 662} (2008) 178;
  H.~Ishimori, T.~Kobayashi, H.~Ohki, Y.~Omura, R.~Takahashi and M.~Tanimoto,
  Phys.\ Rev.\  D {\bf 77} (2008) 115005;
  P.~H.~Frampton and S.~Matsuzaki,
  arXiv:0806.4592 [hep-ph];
  F.~Feruglio, C.~Hagedorn, Y.~Lin and L.~Merlo,
  Nucl.\ Phys.\  B {\bf 809} (2009) 218;
  H.~Ishimori, T.~Kobayashi, Y.~Omura and M.~Tanimoto,
  JHEP {\bf 0812} (2008) 082
  H.~Ishimori, T.~Kobayashi, H.~Okada, Y.~Shimizu and M.~Tanimoto,
  JHEP {\bf 0904} (2009) 011;
  A.~Adulpravitchai, A.~Blum and C.~Hagedorn,
  JHEP {\bf 0903} (2009) 046
  A.~Blum and C.~Hagedorn,
  Nucl.\ Phys.\  B {\bf 821} (2009) 327;
  A.~Hayakawa, H.~Ishimori, Y.~Shimizu and M.~Tanimoto,
  Phys.\ Lett.\  B {\bf 680} (2009) 334;
  H.~Ishimori, T.~Kobayashi, H.~Okada, Y.~Shimizu and M.~Tanimoto,
  arXiv:0907.2006 [hep-ph];
  Y.~Lin, L.~Merlo and A.~Paris,
  arXiv:0911.3037 [hep-ph].

\bibitem{Georgi:2000ks}
  H.~Georgi, A.~K.~Grant and G.~Hailu,
  Phys.\ Lett.\  B {\bf 506} (2001) 207
  [arXiv:hep-ph/0012379].
\end{thebibliography}
\end{document}